\documentclass[sigconf, screen]{acmart}
\usepackage{csquotes}
\MakeOuterQuote{"}
\usepackage{makecell}
\usepackage{tabularx}
\usepackage{booktabs}
\AtBeginDocument{%
  \providecommand\BibTeX{{%
    \normalfont B\kern-0.5em{\scshape i\kern-0.25em b}\kern-0.8em\TeX}}}

\copyrightyear{2023}
\acmYear{2023}
\setcopyright{rightsretained}
\acmConference[UIST '23]{The 36th Annual ACM Symposium on User Interface Software and Technology}{October 29-November 1, 2023}{San Francisco, CA, USA}
\acmBooktitle{The 36th Annual ACM Symposium on User Interface Software and Technology (UIST '23), October 29-November 1, 2023, San Francisco, CA, USA}
\acmDOI{10.1145/3586183.3606823}
\acmISBN{979-8-4007-0132-0/23/10}

\newcommand{\rev}[1] {{#1}}
\begin{document}

\title{Soundify: Matching Sound Effects to Video}

\author{David Chuan-En Lin}
\affiliation{%
  \institution{Carnegie Mellon University}
  \streetaddress{5000 Forbes Ave.}
  \city{Pittsburgh, PA}
  \country{USA}
  }
\email{chuanenl@cs.cmu.edu}

\author{Anastasis Germanidis}
\affiliation{%
  \institution{Runway}
  \streetaddress{}
  \city{New York, NY}
  \country{USA}
  }
\email{anastasis@runwayml.com}

\author{Cristóbal Valenzuela}
\affiliation{%
  \institution{Runway}
  \streetaddress{}
  \city{New York, NY}
  \country{USA}
  }
\email{cris@runwayml.com}

\author{Yining Shi}
\affiliation{%
  \institution{Runway}
  \streetaddress{}
  \city{New York, NY}
  \country{USA}
  }
\email{yining@runwayml.com}

\author{Nikolas Martelaro}
\affiliation{%
  \institution{Carnegie Mellon University}
  \streetaddress{5000 Forbes Ave.}
  \city{Pittsburgh, PA}
  \country{USA}
  }
\email{nikmart@cmu.edu}

\renewcommand{\shortauthors}{David Chuan-En Lin, Anastasis Germanidis, Cristóbal Valenzuela, Yining Shi, and Nikolas Martelaro}

\begin{abstract}
  In the art of video editing, sound helps add character to an object and immerse the viewer within a space. Through formative interviews with professional editors ($N$=10), we found that the task of adding sounds to video can be challenging. This paper presents Soundify, a system that assists editors in matching sounds to video. Given a video, Soundify identifies matching sounds, synchronizes the sounds to the video, and dynamically adjusts panning and volume to create spatial audio. In a human evaluation study ($N$=889), we show that Soundify is capable of matching sounds to video out-of-the-box for a diverse range of audio categories. In a within-subjects expert study ($N$=12), we demonstrate the usefulness of Soundify in helping video editors match sounds to video with lighter workload, reduced task completion time, and improved usability.
\end{abstract}



\begin{teaserfigure}
  \centering
  \includegraphics[width=15cm]{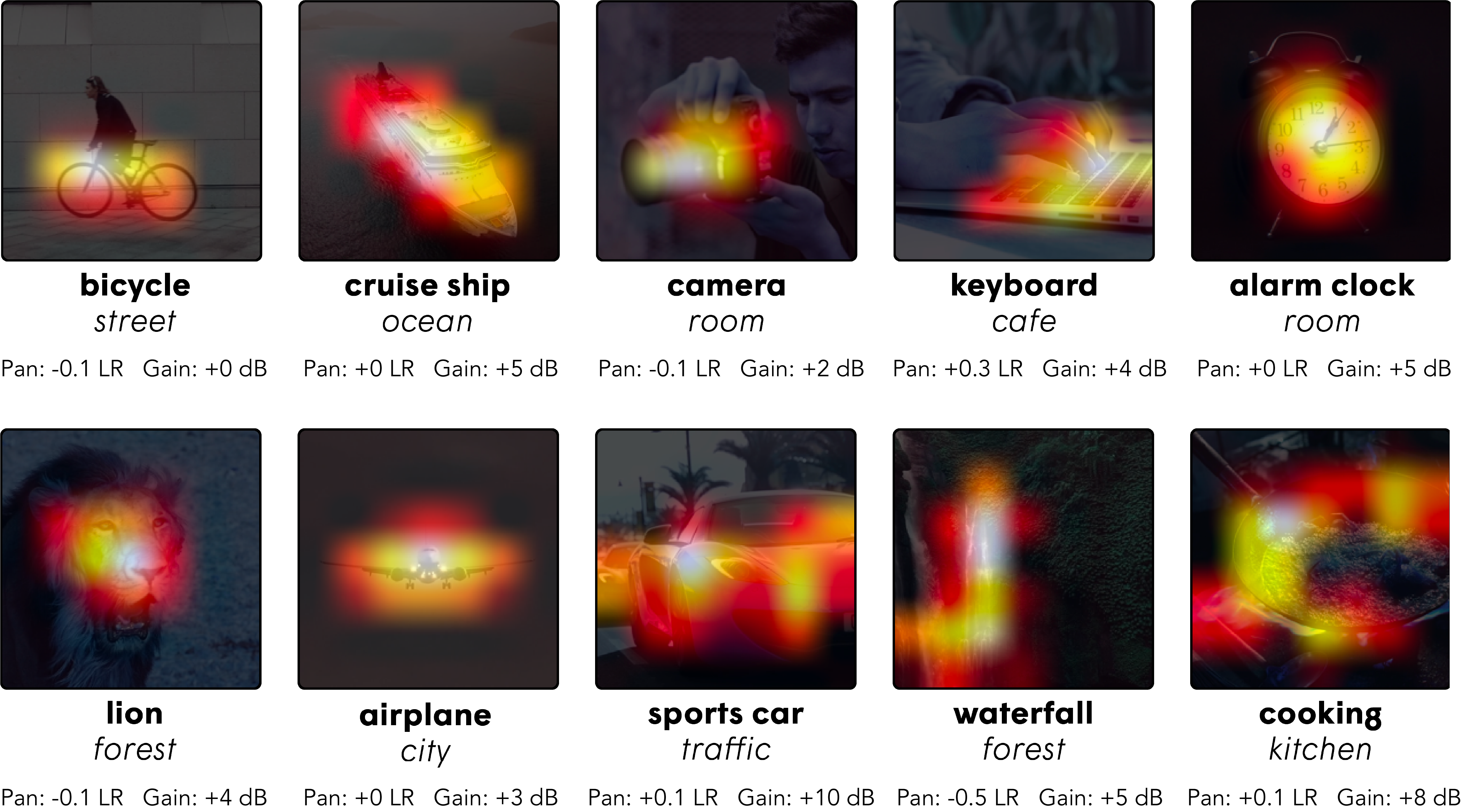}
  \caption{Soundify assists users in matching sound effects (in bold) and ambients (in italics) to video, and helps dynamically adjust panning and volume by localizing "sound emitters."}
  \Description{Soundify assists users in matching sound effects (in bold) and ambients (in italics) to video, and helps dynamically adjust panning and volume by localizing "sound emitters."}
  \label{fig:teaser}
\end{teaserfigure}

\maketitle

\section{Introduction}

\vspace{0.3cm}
\begin{quotation}
\textit{``Sound is half the experience in seeing a film.''}
\vspace{0.1cm}
\par\raggedleft--- \textup{George Lucas, Film Director}
\end{quotation}
\vspace{0.3cm}

In the art of video editing, sound helps add character to an object and immerse the viewer within a space. Although a video's soundscape may be built by sounds recorded on set, it is common for editors to replace or add complementary sounds from external sound collections. The process of adding sounds to video is called the \textit{foley pass} \cite{foley}. During the foley pass, a skilled video editor analyzes the scene and overlays sounds, such as \textit{effects} (e.g., bicycle bell ring) and \textit{ambients} (e.g., street noise). Through formative interviews with 10 professional video editors, we found that this process can be challenging and time-consuming, especially as the amount of video footage scales up. We thus aim to develop a system to assist editors in adding sound to videos.

Many approaches have been proposed to synthesize audio for videos. For example, Visual to Sound \cite{zhou2018visual} learns to generate waveform audio given a visual input and AutoFoley \cite{ghose2020autofoley} learns to synthesize foley sounds for video clips.
However, these works necessitate extensive training of audio generation models over large datasets and often produce subpar sounds containing undesirable noise and artifacts.
Our work takes a different approach by leveraging existing studio-quality sound effects libraries. These libraries contain sound clips recorded by experts using first-rate equipment under pristine conditions.
Rather than synthesizing audio from scratch, we investigate the alternative approach of retrieving matching high-quality sound clips, then remixing them (i.e., pan and volume parameters) to fit the video footage (Figure \ref{fig:pan-gain}).
Our key insight is to extend CLIP \cite{radford2021learning}, an image classification neural network that has learned a joint-embedding space between image and text, into a "zero-shot sound detector" through probing the model's activation maps (Figure \ref{fig:teaser}). Given a library of sound clips, we compare the labels of the sound clips to the video to identify the sound(s) that are present in the video. Next, we tune the left-right panning according to the activation map coordinates and the volume according to the activation map area.

In this paper, we present Soundify, a system that assists video editors in matching sound effects to video. From our interviews with professional video editors, we distill four key design principles for Soundify to assist editors---surface, synchronize, spatial, and stack. Given video footage, Soundify helps the video editor \textit{surface} matching sound clips, \textit{synchronize} the sound clips to the objects in the video, dynamically adjusts the \textit{spatial} aspect of the sound clips (panning and volume) based on the video content, and allows the \textit{stacking} of multiple sound clips such as foreground objects and ambients to create an immersive soundscape. We test the capability of Soundify against a baseline by matching a wide variety of sounds to complex videos and collecting human ratings from Mechanical Turk. Then, we evaluate the usefulness of Soundify in an expert study with video editors against a baseline, demonstrating improvements in workload, usability, and task completion time. We encourage you to have a look at, or better yet, have a \textit{listen} to the results of Soundify at \url{https://chuanenlin.com/soundify}.


In summary, this paper makes the following contributions:
\begin{itemize}
\item {Soundify, a system that assists video editors in matching sounds to video.}
\item{A human evaluation ($N$=889 raters) testing the functionality of Soundify against a baseline. Participants report a significant preference for Soundify's results.}
\item{Insights from an expert study ($N$=12 professional video editors) with Soundify against a manual editing baseline. Participants experience lighter workload, lower task completion time, and higher usability.}
\end{itemize}

\section{Related Work}
Our work is situated among literature in two main branches of research in matching sounds to video: audio synthesis and audio-visual correspondence learning. 
Further, we refer to prior HCI works on systems to assist users with audio editing.

\subsection{Audio Synthesis}
Audio synthesis is an actively explored topic in the audio research community. Research in this space typically adopts generative models to synthesize raw audio. \cite{oord2016wavenet} introduces a deep neural network for synthesizing waveform speech from scratch by training on tens of thousands of audio samples in an autoregressive manner. \cite{engel2019gansynth, kumar2019melgan} learn to synthesize coherent waveform audio with Generative Adversarial Networks (GANs). Rather than synthesizing waveform audio, \cite{yang2017midinet} trains a GAN to generate MIDI music notes. \cite{dhariwal2020jukebox, kong2020diffwave} explore other types of generative architectures for raw audio generation, including a multi-scale VQ-VAE and a diffusion-based model.

Among works in audio synthesis, several works investigate generating audio specifically based on visual input. \cite{zhou2018visual, chen2020generating, owens2016visually, chen2018visually} explore generating raw audio based on video frames. \cite{gan2020foley, su2020audeo} introduce systems that synthesize plausible music based on videos of people performing musical instruments. \cite{su2021does, aggarwal2021dance2music} learn to generate music driven by human skeleton key points. More recently, \cite{ghose2020autofoley, ghose2021foleygan} learn to synthesize foley-like audio tracks from videos by training on a combination of self-recorded foley clips and YouTube videos. Nonetheless, generative audio synthesis approaches require large amounts of curated data. In addition, generated waveform audio results may contain noise artifacts. On the other hand, professional video editing workflows have high sound quality requirements and typically use clean, studio-recorded sounds. In this work, rather than generating raw audio from scratch, we retrieve audio clips from high-quality audio libraries, then tune the panning and volume of the audio based on the video.

\subsection{Audio-Visual Correspondence Learning}
More recently, researchers have investigated the task of learning audio-visual correspondence (i.e., learning the association between images and audio), typically from large labeled datasets. The audio-visual correspondence learning task was first defined by \cite{arandjelovic2017look} where the authors train separate visual and audio networks with simple late fusion layers to determine whether an audio-visual pair has correspondence or not. Newer works learn audio-visual correspondence at scale by exploiting the natural synchronization of video frames and audio present in web videos \cite{senocak2018learning, zhao2018sound, gao2019co, tian2018audio, gao2018learning}. As audio-visual pairs harvested from videos are noisy in terms of quality, researchers often adopt a weakly-supervised training approach with massive amounts of videos (e.g., several thousands to millions of videos). In this work, we bypass the need to learn audio-visual correspondence from large-scale data by leveraging libraries of labeled audio clips and utilizing the image-text correspondence capability of CLIP \cite{radford2021learning}, a neural network trained on web-scale image-caption pairs. We further extend CLIP to support the fine-grained localization of sound emitters by exploiting activation maps \cite{selvaraju2016grad}, which models where the neural network is "looking at", to achieve spatial audio.

\subsection{Audio-Related Multimedia Editing in HCI}
\rev{
HCI researchers have developed many tools for audio-related multimedia editing. Soundify builds on a large thread of work that leverages algorithmic and AI techniques to help users complete creative tasks more easily and effectively.
\cite{xia2020crosscast} helps users add visuals to travel podcasts by mining geographic locations and descriptive keywords. \cite{xia2020crosspower} proposes interaction techniques that enable users to add graphical content based on a script or outline. \cite{shin2016dynamic} introduces a workflow that combines the editing processes of script writing and audio recording for creating podcasts. \cite{wang2022record} helps users automatically shorten voice recordings given a length constraint to cater to different social media platforms. \cite{rubin2013content} assists users in composing audio stories with a transcript-based speech editing tool and an emotion-driven music browser. \cite{rubin2014generating} helps users generate emotionally relevant music scores for audio stories. \cite{chong2023soundtoons} enables novice creators to author live audio-driven animations. \cite{smaragdis2009user} presents an interface for isolating and selecting specific sounds within audio mixtures.
This paper aims to contribute to this thread of work focused on developing systems that augment the capabilities of creators.
While many current systems primarily assist users by analyzing textual content, such as scripts, we explore an approach that focuses on analyzing the visuals rather than the text of the content.
}

\begin{table*}[]
\begin{tabular}{@{}lcp{2in}p{3.7in}@{}}
\toprule
\textbf{ID} &
  \multicolumn{1}{l}{\textbf{Experience}} &
  \textbf{Video and Audio Editing Tool(s)} &
  \textbf{Short biography} \\ \midrule
I1 &
  20 years &
  Adobe Premiere Pro, Avid Pro Tools &
  Filmmaker, producer, works on feature films, animation, and virtual reality \\ \midrule
I2 &
  15 years &
  Adobe Premiere Pro, Avid Pro Tools &
  Filmmaker, producer, works on feature films, documentaries, and interactive media \\ \midrule
I3 &
  13 years &
  Adobe Premiere Pro, DaVinci Resolve &
  Video editor, VFX artist, director, owns a production studio, YouTuber (9M subs) \\ \midrule
I4 & 11 years & Adobe Premiere Pro, Adobe After Effects, Timebolt & Video editor, VFX artist, worked in an advertising agency, YouTuber (300K subs)              \\ \midrule
I5 & 12 years & Adobe Premiere Pro, Adobe Audition                & Video editor, VFX artist, worked in software engineering, Instagram creator (400K followers) \\ \midrule
I6 &
  15 years &
  Adobe After Effects &
  Video editor, VFX artist, animator, skateboarder, worked with advertising agencies and athletes \\ \midrule
I7 & 20 years & Adobe Premiere Pro, Final Cut Pro                 & Video editor, creative director, worked on AAA video game videos and government campaigns    \\ \midrule
I8 &
  8 years &
  Final Cut Pro, iMovie &
  Video editor, comedian, works on commentaries, former Viner, YouTuber (4M subs) \\ \midrule
I9 &
  9 years &
  Adobe Premiere Pro &
  Video editor, animator, programmer, works on tech videos and tutorials, YouTuber (50K subs) \\ \midrule
I10 &
  11 years &
  Avid Pro Tools &
  Audio editor, programmer, worked in video game sound design and at an AI research lab \\ \bottomrule
\end{tabular}
\label{table:interviews}
\caption{Interviewee information. We assign an ID to each interviewee (later referenced in Section \ref{section:principles}). We list each interviewee's years of professional video and audio editing experience, their preferred suite of editing tools, and a short biography.}
\end{table*}

\section{Formative Interviews}
To formulate the design principles for Soundify, we conduct a set of formative interviews with 10 professional video editors to understand their creative processes and uncover potential areas for improvement in their sound editing work.
Below, we describe our participants and interview procedure.
We then describe the findings from the interviews as part of our discussion on the Design Principles in Section \ref{section:principles}.

\subsection{Participants}

We recruited 10 professional video editors (I1 - I10, 2 female, 8 male) from known contacts and social media postings. Our interviewees have varying years of video and audio editing experience (mean=13.80, SD=4.29), work in a variety of roles (e.g., filmmaker, visual effects artist, comedian, programmer, content creator), and work with various types of video (e.g., feature films, commercials, documentaries, animation, online videos) (Table 1). All of our participants actively edit video for their work and add sound effects to enhance the content that they create.

\subsection{Procedure}

We conduct our interviews over video conference. Our interviews consist of three stages. First, we ask the participant to provide an overview of their video and audio editing experience. Second, we ask the participant to describe their typical workflow of adding sounds to video. Finally, we ask the participant to reflect on potential features they would like to see in future video and audio editing tools. We record the interviews and also take notes throughout the interviews. The interview lasts for approximately 30 minutes.

\begin{figure}[tbp]
  \includegraphics[width=\columnwidth]{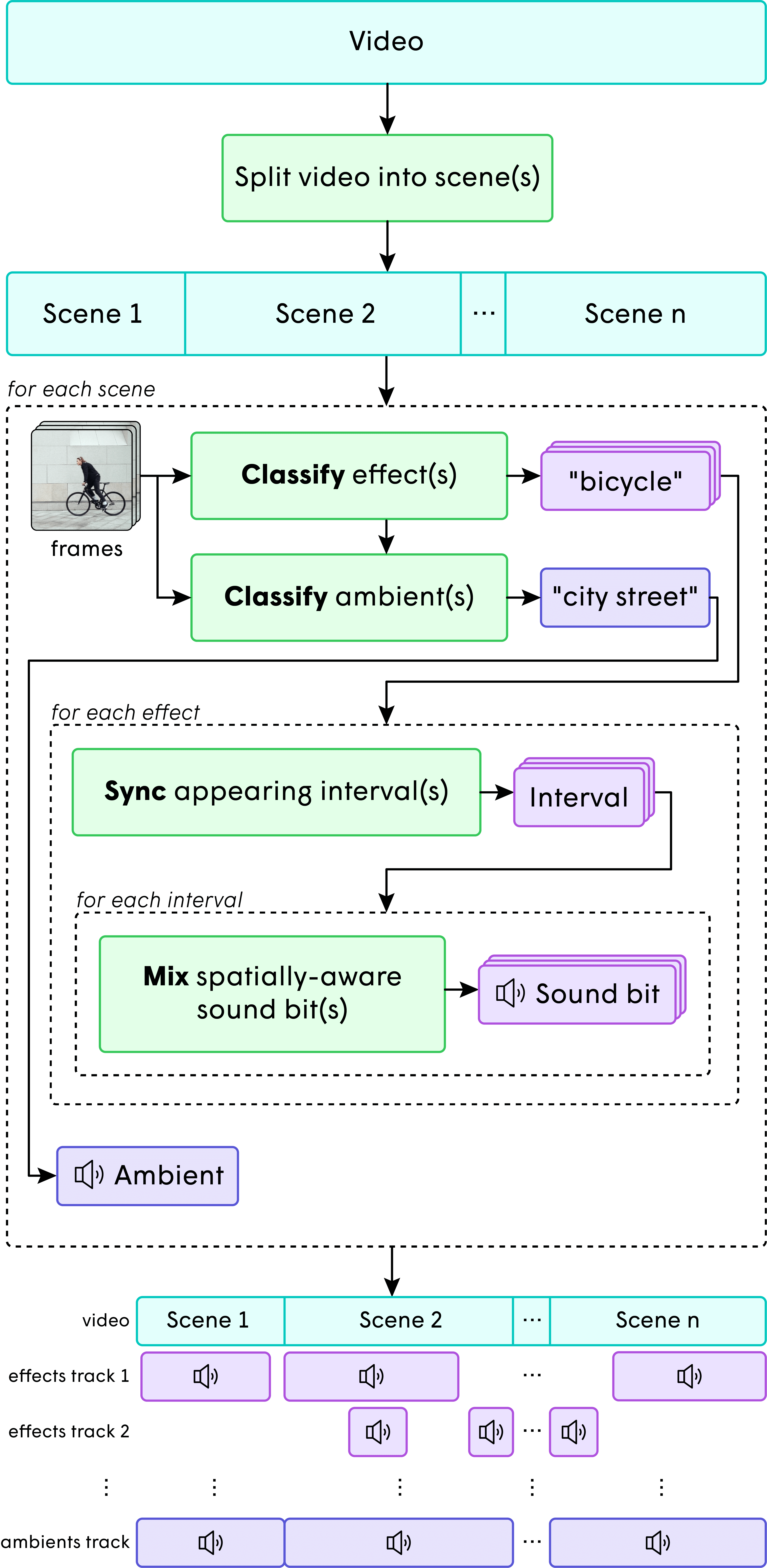}
  \caption{Overview of Soundify. Soundify first splits a video into scenes. For each scene, Soundify classifies for effects and ambients. The matched ambient is used for the entire scene. For each matched effect, Soundify performs more fine-grained synchronization by identifying their appearing intervals. For each interval, Soundify mixes spatial sound chunks with computed pan and gain parameters. The final result consists of one or more effects tracks and an ambients track.}
  \Description{Overview of Soundify. Soundify first splits a video into scenes. For each scene, Soundify classifies for effects and ambients. The matched ambient is used for the entire scene. For each matched effect, Soundify performs more fine-grained synchronization by identifying their appearing intervals. For each interval, Soundify mixes spatial sound chunks with computed pan and gain parameters. The final result consists of one or more effects tracks and an ambients track.}
  \label{fig:overview}
\end{figure}

\section{Principles of Sound to Video Matching}
\label{section:principles}
We analyze our formative interviews with inductive thematic analysis \cite{thematic-analysis}. We grouped interviewees' quotes into a set of themes, which became our four key principles for the task of matching sound to video. These principles guide the development of Soundify.

\subsection{Principle 1: Surface}
First, Soundify needs to help editors surface relevant audio clips based on the video content. Most of the editors we interviewed frequently use libraries of high-quality sounds, such as \href{https://www.epidemicsound.com/}{Epidemic Sound}, in their sound editing workflows. While it may not seem like much effort to find a sound with keyword searching and add it to a video, the effort piles up with many full-length videos containing hundreds or thousands of video clips and numerous sounds per clip. In addition, editors often see finding and adding background sounds to video as a non-creative task: "\textit{I just want it to sound like a forest… it's just a task I want to do along the way.} (I3)" Surfacing relevant sounds could potentially speed up an editor's workflow considerably.

\subsection{Principle 2: Synchronize}
Second, Soundify needs to help editors synchronize the surfaced audio clips based on the video clip. For example, given the sound effect of a bicycle peddling, the audio needs to come in when the bicycle appears on the scene and go away when the bicycle exits the scene. Nonetheless, given an audio clip, it can be tedious to manually align it to select video frames. Editors expressed that synchronizing sound can be a "\textit{laborious process} (I10)" since it involves a lot of trial and error and suggested that "\textit{a lot can be done in automated syncing} (I10)".

\begin{figure*}[tbp]
  \centering
  \includegraphics[width=13cm]{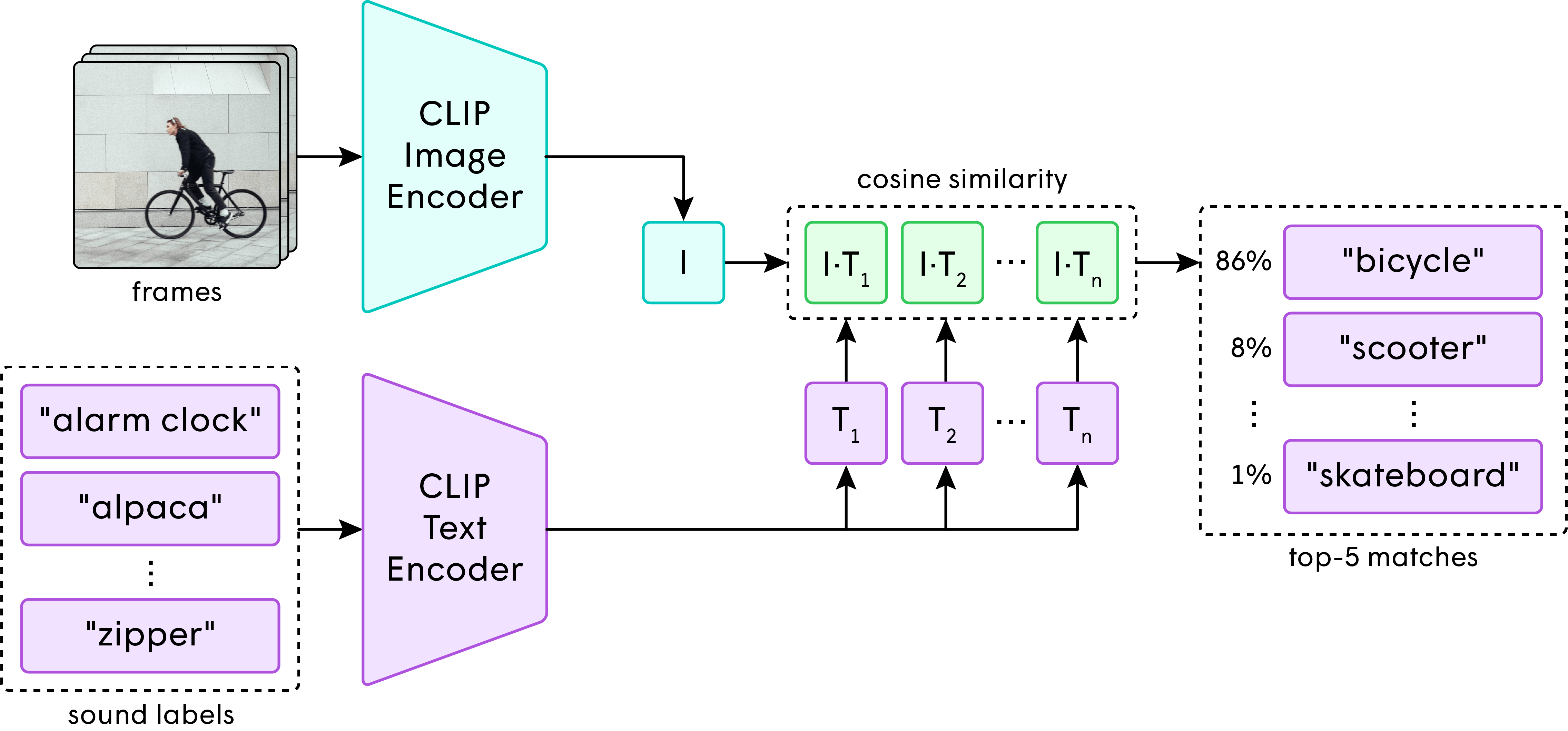}
  \caption{Effects classification. Given the frames of a scene and a database of sound labels, Soundify performs pairwise comparisons to predict the top-5 matching sounds.}
  \Description{Effects classification. Given the frames of a scene and a database of sound labels, Soundify performs pairwise comparisons to predict the top-5 matching sounds.}
  \label{fig:classify}
\end{figure*}

\begin{figure*}[tbp]
  \centering
  \includegraphics[width=13cm]{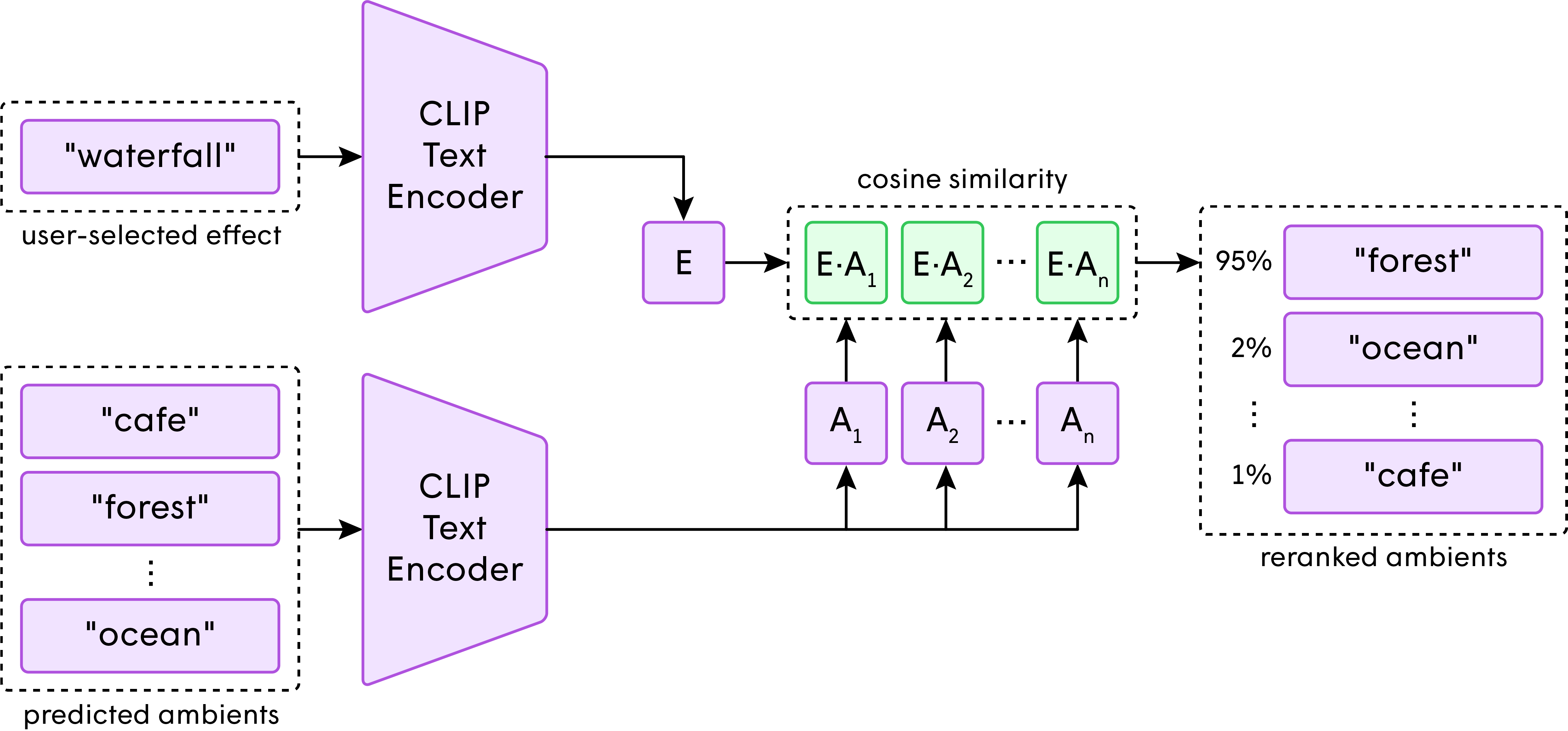}
  \caption{Ambients classification. Since ambients classification can be more error-prone, given the user-select effects label and predicted ambients labels, Soundify performs pairwise comparisons to rerank the ambients.}
  \Description{Ambients classification. Since ambients classification can be more error-prone, given the user-select effects label and predicted ambients labels, Soundify performs pairwise comparisons to rerank the ambients.}
  \label{fig:ambients}
\end{figure*}

\subsection{Principle 3: Spatial}
Third, Soundify needs to help editors dynamically convert the audio clip to spatial sound. Using the bicycle example, as the bicycle peddles from left to right, the audio clip should gradually move from the audience's left ear to the right ear (i.e., audio panning). If the bicycle starts off far away and moves closer to the viewer, the audio clip should gradually become louder over time (i.e., audio volume). However, similar to synchronizing, tuning pan and volume parameters of an audio frame by frame can become "\textit{the tedious part that no one wants to do} (I5)". Editors hoped there could be "\textit{a clever way of generating stereo} (I3)" since it can be "\textit{one of the most difficult and time-consuming parts} (I3)".

\subsection{Principle 4: Stack}
Fourth, Soundify needs to give editors the ability to stack multiple audio tracks. Editors expressed that their sound editing workflow involves "\textit{a lot of blending and layering} (I9)". Similar to how choirs consist of multiple singers singing at different ranges to create a sound that is fuller and has more depth, a good soundscape involves the stacking of multiple audio tracks. Audio stacking usually involves a base layer of \textit{ambients} and several layers of \textit{effects}. For example, an editor describes his workflow for stacking audio as: "\textit{Ambient room tones in one track, voice-over on another track, and six tracks for effects} (I1)". An editor further emphasized the importance of having ambient noise even if there are no clear effects to be added: "\textit{It's really important to always have `something'... even some ambient street sound to make it sound natural… it is always better to have something than nothing} (I3)".

\begin{figure*}[tbp]
  \centering
  \includegraphics[width=13cm]{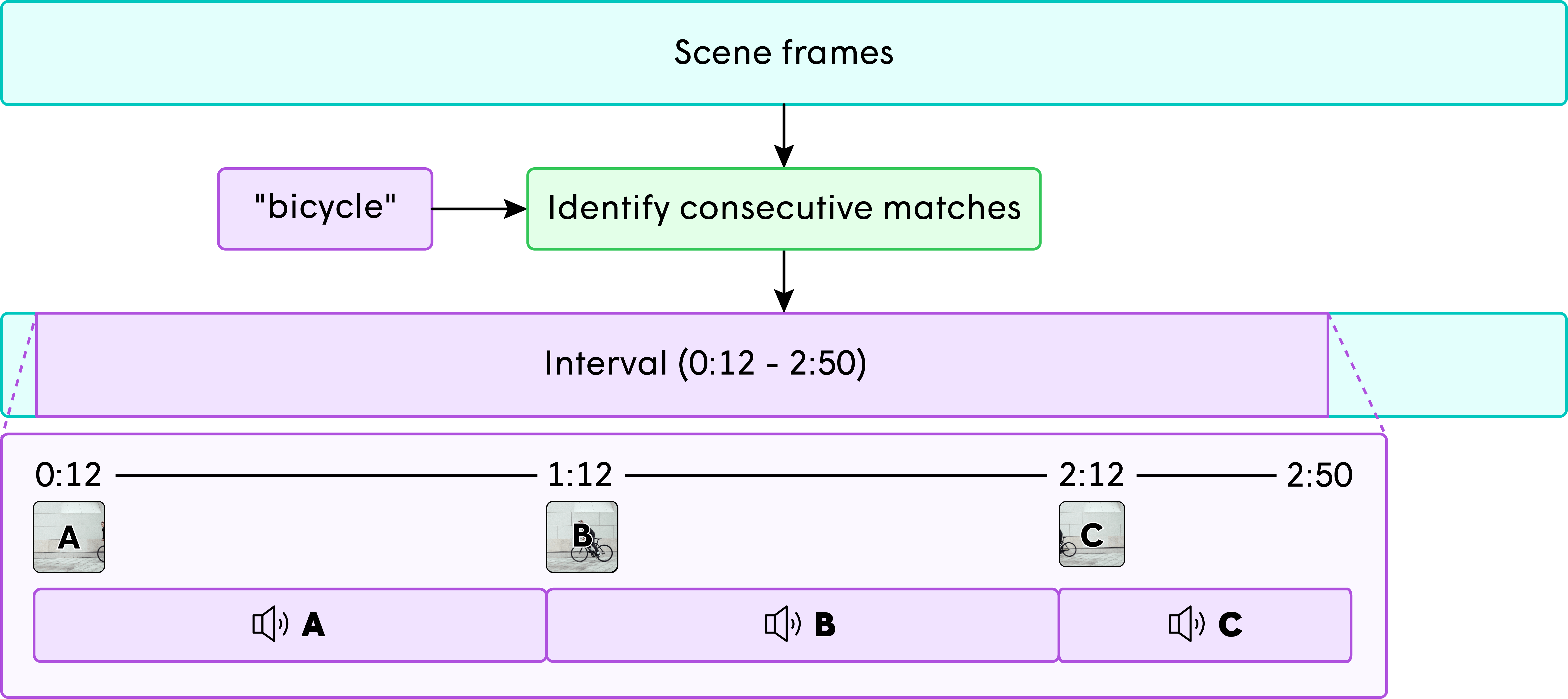}
  \caption{Sync. Given the frames of a scene and a sound label, Soundify identifies appearing intervals. An interval is split into chunks. Each chunk takes the first frame as its reference frame.}
  \Description{Sync. Given the frames of a scene and a sound label, Soundify identifies appearing intervals. An interval is split into chunks. Each chunk takes the first frame as its reference frame.}
  \label{fig:sync}
\end{figure*}

\begin{figure*}[tbp]
  \centering
  \includegraphics[width=13cm]{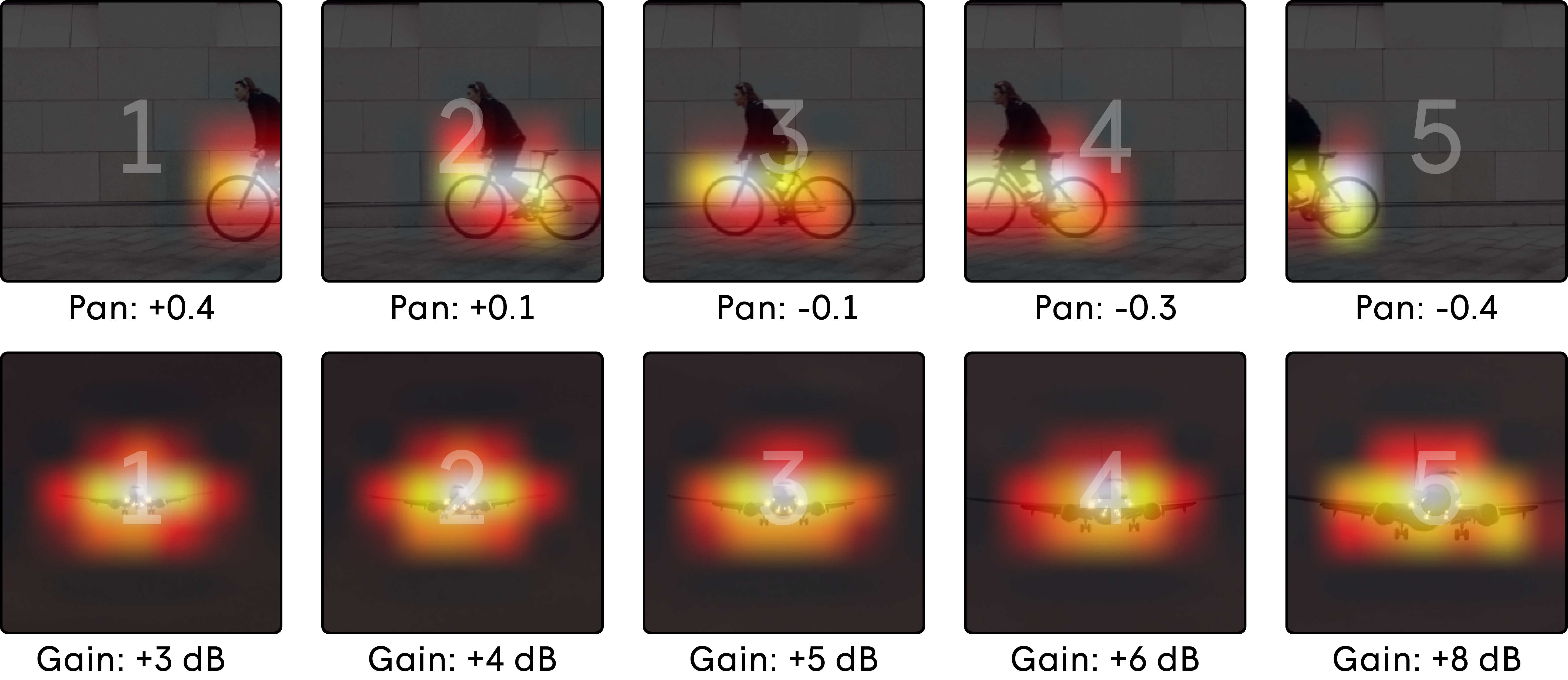}
  \caption{Soundify adapts pan (top row) and gain (bottom row) parameters over time based on the heatmap's position and size.}
  \Description{Soundify adapts pan (top row) and gain (bottom row) parameters over time based on the heatmap's position and size}
  \label{fig:pan-gain}
\end{figure*}

\begin{figure*}[tbp]
  \centering
  \includegraphics[width=13cm]{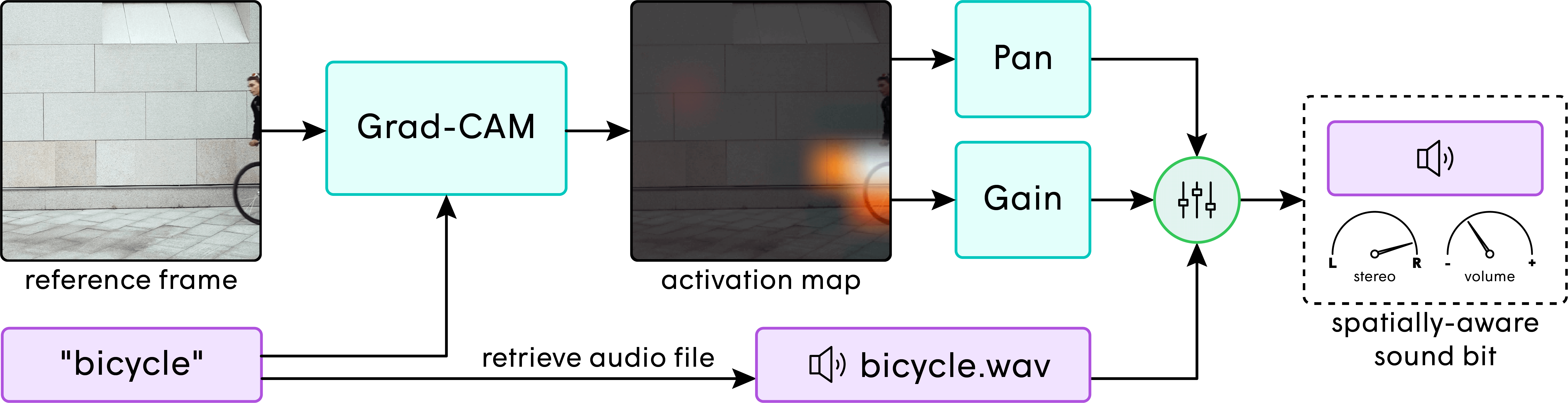}
  \caption{Mix. Given a reference frame and a sound label, Soundify retrieves the relevant audio file and mixes its pan and gain parameters, by referencing the activation map, to generate a spatial sound chunk.}
  \Description{Mix. Given a reference frame and a sound label, Soundify retrieves the relevant audio file and mixes its pan and gain parameters, by referencing the activation map, to generate a spatial sound chunk.}
  \label{fig:mix}
\end{figure*}

\section{Implementation}
\label{section:implementation}
Our four principles are manifested in Soundify and guide its implementation. We first give an overview of our system (Figure \ref{fig:overview}).
To reduce the number of distinct sounds to classify across a long video, we split a given video into scenes. We use a boundary detection algorithm based on color histogram distances \cite{hafner1995efficient}. A large distance between the histograms of neighboring frames indicates a scene change. For each scene, we classify for multiple sound effects and an ambient. For each classified effect, we sync the audio clip to the interval in which the corresponding object appears in the video. The classified ambient is used for the entire scene. For each synchronized interval, we mix the pan and volume parameters of the sound effects over time based on the object's position and size in the video. Finally, we stack our matched effects and ambient to produce the final result. In the following sections, we describe in more detail the implementation of the various components including (1) Classify, (2) Sync, and (3) Mix.

\begin{figure}[tbp]
  \centering
  \includegraphics[width=\columnwidth]{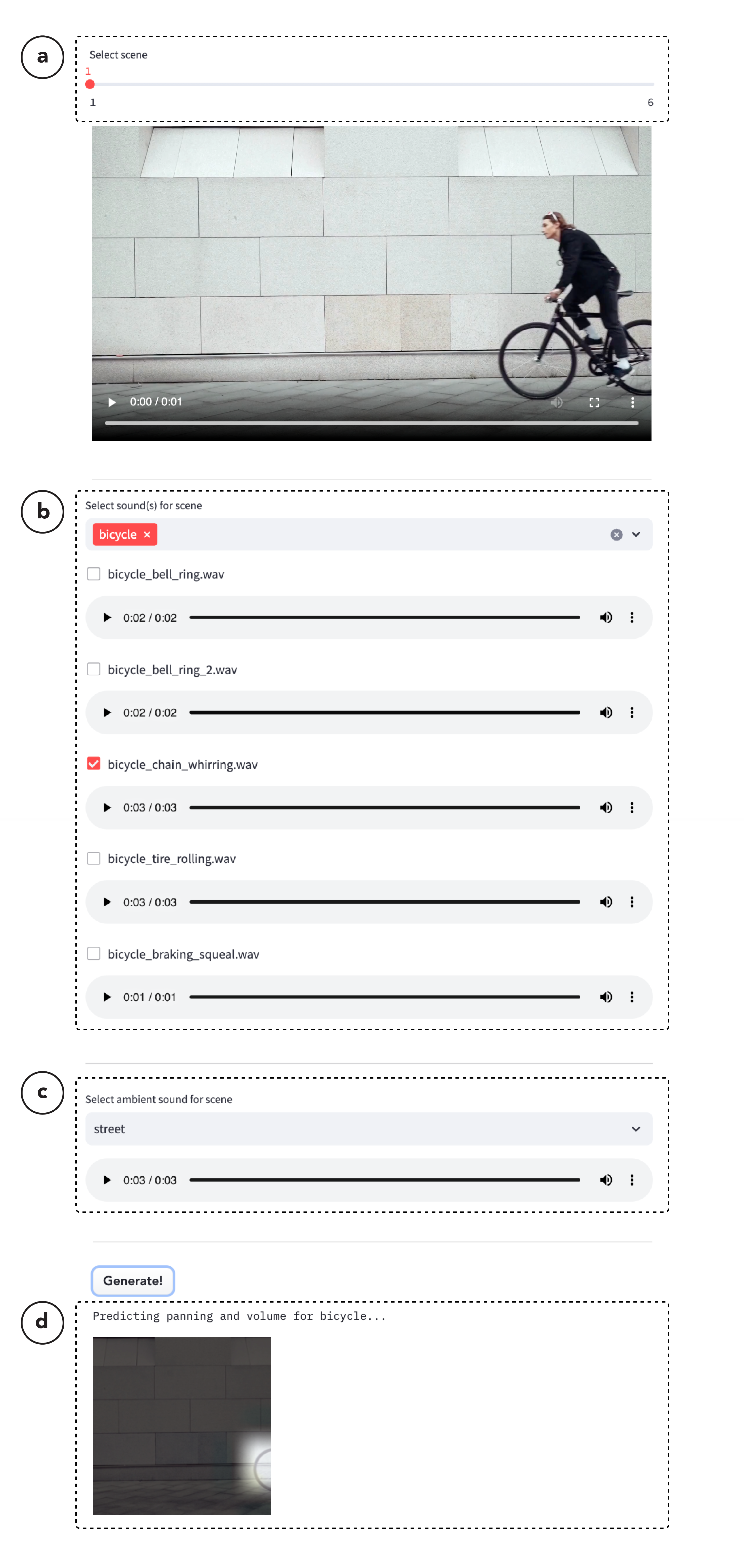}
  \caption{Soundify interface. A video is split into scenes and the user may switch between scenes via a slider (a). For each scene, the user may choose one or multiple recommended sound effects (b) and a recommended ambient (c). After the user hits the "Generate" button, Soundify locates the sounds in the video and the interface displays the predictions for pan and gain levels in one-second intervals (d).}
  \Description{Soundify interface. A video is split into scenes and the user may switch between scenes via a slider (a). For each scene, the user may choose one or multiple recommended sound effects (b) and a recommended ambient (c). After the user hits the "Generate" button, Soundify locates the sounds in the video and the interface displays the predictions for pan and gain levels in one-second intervals (d).}
  \label{fig:interface}
\end{figure}

\subsection{Classify}
\label{classify}
The first stage of Soundify is classification. We match sound effects to a video by classifying "sound emitters" within the video (Figure \ref{fig:classify}). A sound emitter is simply an object or environment that produces sound and is defined based on the sound categories of Epidemic Sound \cite{epidemic-sound}, a curated database of over 90,000 high-quality sound effects. To construct a realistic soundscape, we classify each scene for two types of sounds: \textit{effects} (e.g., bicycle, camera, keyboard) and \textit{ambients} (e.g., street, room, cafe) (Principle 1). For a given scene, we run each video frame through the CLIP image encoder and concatenate the encoded frames into a vector representation for the entire scene. For each \textit{effects} label in the sound database, we run it through the CLIP text encoder to return a vector representation for the label. We then perform pairwise comparisons between the encoded scene vector and each encoded effects label vector with cosine similarities and obtain the top-5 matching effects labels for the scene. The user may then select one or more recommended effects (Principle 4). For \textit{ambients} type labels, we perform the same encoding and pairwise comparison steps. Ambients classification can be more error-prone due to the background potentially being visually out of focus. Thus, we additionally run both the predicted ambients and the previously \textit{user-selected} effect(s) through CLIP text encoders, and rerank the predicted ambients based on their cosine similarities (Figure \ref{fig:ambients}). For example, \verb|forest| may be ranked higher than \verb|cafe| if the user had previously selected \verb|waterfall| as the effect. The user may then select a recommended ambient.

\subsection{Sync}
A sound emitter may appear on screen for only a subset of the scene. Therefore, we want to synchronize effects to when their sound emitter appears (Principle 2) (Figure \ref{fig:sync}). We pinpoint such intervals by comparing the effects label with each frame of the scene. If a sequence of consecutive frames have similarity scores above a threshold, we identify it as an interval. There may be multiple intervals in each scene, such as when a sound emitter disappears then reappears.

\subsection{Mix}
Video editors adjust sound according to the state of the scene. For instance, as a \verb|bicycle| peddles from one side to another, we hear a shift in stereo panning (i.e., sound moves from left to right). As an \verb|airplane| glides up close, we experience a gain in sound intensity (i.e., sound volume changes). Similarly, we mix an effect's pan and gain parameters over time (Principle 3) (Figure \ref{fig:pan-gain}). To achieve this, we split an effects interval into around one-second chunks (Figure \ref{fig:sync}), mix the pan and gain parameters for each chunk (Figure \ref{fig:mix}), and stitch the chunks smoothly with crossfades. A one-second chunk uses the first image frame as the reference image. We run the reference image through Grad-CAM \cite{selvaraju2016grad} on the ReLU activation of the last visual layer (ResNet-50 architecture) to generate an activation map (example visualizations of activation maps shown in Figure \ref{fig:pan-gain}). This \textit{localizes} the sound emitter, allowing the system to take on the capabilities of an \textit{open-vocabulary} object detector (i.e., works on arbitrary objects). We then compute the pan parameter by the x-axis of the localized sound emitter's center of mass and the gain parameter by its normalized area. Next, we retrieve the effect's corresponding \verb|.wav| audio file and remix its pan and gain. \rev{We prioritize retrieving audio clips that have a duration longer or equal to the occurrence of a sound object on screen.} For ambients, we assume a constant environment for each scene. Thus, we retrieve the corresponding \verb|.wav| audio file and use it across the entire scene with a -5 dB volume adjustment as to not overpower the main sound effects. Finally, we stack all selected audio tracks of effects and ambients for all scenes into one final audio track for the video (Principle 4) (Figure \ref{fig:overview}).

\subsection{Interface}
We provide users with an interface to view the system's predictions and make creative sound design decisions (Figure \ref{fig:interface}). We first split a video into various scenes and allow users to adjust the sound effects and ambients for each scene. The user may switch between scenes using a slider (Figure \ref{fig:interface}a). In the sound effects panel (Figure \ref{fig:interface}b), the highest-scoring sound effect is pre-populated by default. The user may add or remove sound effects and stack multiple sound effects (Principle 4) from a multi-selection dropdown menu with sound effects sorted in descending order of their predicted scores. The user may preview the audio file of each selected sound effect. In the ambients panel (Figure \ref{fig:interface}c), the highest-scoring ambient is pre-populated by default. The user may switch to another ambient from a dropdown menu with ambients sorted in descending order of their predicted scores. The user may also preview the audio file of the selected ambient. After the user clicks on the "Generate" button, we visualize the heatmap predictions in one-second intervals (Figure \ref{fig:interface}d). After the generation is complete, we also allow users to export the video and audio tracks of the video split and numbered by scene (i.e., scene 1 video track, scene 1 audio track, …) as a \texttt{.zip} file so that they may import and use the audio clips in a video editor of their choice.

\section{Human Evaluation}

To test the effectiveness of Soundify in detecting sound emitters and matching sounds to them, we run a human evaluation study on a collection of videos with sounds matched by Soundify against a baseline method built on YOLO \cite{yolo}, \rev{a state-of-the-art object detector trained on 328,000 images of 91 categories of everyday objects and humans, many of which can be found in the videos we are testing. YOLO provides a bounding box around objects that can allow for us to track them across frames and adjust pan and volume based on position and size.} We do not compare against an audio synthesis model as our baseline as the focus of this paper is on investigating the utilization of existing sound effects clips. The following outlines our experimental setup, procedure, and results.

\begin{figure}[tbp]
  \centering
  \includegraphics[width=\columnwidth]{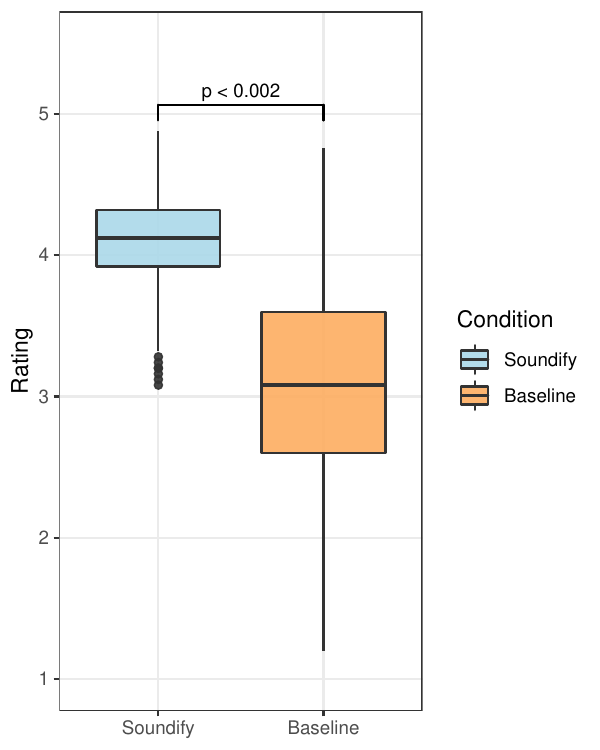}
  \caption{Human evaluation results ($N$=889) (5-point Likert scale, higher is better).}
  \Description{Human evaluation results (N=889) (5-point Likert scale, higher is better).}
  \label{fig:human-evaluation-results}
\end{figure}

\subsection{Setup}

\subsubsection{Source Material Collection}

We first collect our set of audio clips and video clips that we would like to match together.

\textbf{Audio Clip Collection.} For our audio clips, we use a combination of the ESC-50 dataset \cite{piczak2015esc} and a subset of the UrbanSound8K dataset \cite{salamon2014dataset}. In total, we obtain a diverse audio collection of 54 categories with 40 audio clips per category, yielding a set of 2,160 audio clips.

\textbf{Video Clip Collection.} We collect video clips from Getty Images, a large database of professional video footage. We query for videos on Getty Images using the audio category labels as keywords and collect the first 80 results for each keyword, while manually filtering out irrelevant videos. In total, we result in a set of 1,105 videos clips that represent diverse and complex scenes containing multiple objects.

\subsubsection{Soundify and Baseline Setup}

We match audio clips to video, including synchronization and spatial tuning of pan and volume, with two systems: Soundify and the baseline system. \rev{The sound matching processes were done automatically by the two systems.}

\textbf{Soundify Setup.} We match sound to video with Soundify as detailed in Section \ref{section:implementation}.

\textbf{Baseline Setup.} Our baseline system is based on YOLO \cite{yolo}, an object detector that draws bounding boxes around detected objects. We use a YOLO model trained on a large variety of categories in the COCO dataset \cite{lin2014microsoft}. After the model detects the objects, we take the average x-value of the bounding box coordinates to set the sound panning and the area of the boundary box for the sound volume.

\subsection{Procedure}

We run a Mechanical Turk study to evaluate the videos with sound matched by Soundify and the baseline system. We set the worker qualification requirements of above 95\% approval rate and greater than 100 HITs approved. We also ask that workers have headphones with left and right channels to participate in our task. For each video matched with sound, we ask five different people to evaluate it. It takes around 12 seconds to complete a HIT and we pay \$0.04 USD per HIT, which is above US minimum wage. The worker does not know whether a given video is matched by Soundify or the baseline system. We ask workers to answer how they feel about each of the following statements on a scale of 1 to 5 (strongly disagree, disagree, neutral, agree, strongly agree):

\begin{itemize}
\item {The audio matches the type of object shown in the video.}
\item{The audio aligns in time well with the video.}
\item{The volume of the audio matches well with the video.}
\item{The panning of the audio matches well with the video.}
\item{Overall, the audio matches well with the video.}
\end{itemize}

After collecting the responses, we filter out responses where the worker does not play the video or does not take a long enough time to complete the task.

\subsection{Results}

Figure \ref{fig:human-evaluation-results} shows an overview of the human evaluation results comparing Soundify and the baseline system. We analyze the results for statistical significance through an unpaired t-test with Bonferroni correction (5 tests, significance level at $\alpha$<$\frac{0.01}{5}$=0.002). Participants report a significantly higher rating for the results generated with Soundify (mean=4.11, SD=0.29) compared to the baseline (mean=3.09, SD=0.69) ($t$(1136.1)=-40.80, $p$<0.002, $r$=0.77, $d_s$=1.92) (5-point Likert scale).

\rev{Our results suggest how our extension of CLIP with activation maps can enable open-vocabulary object detection, meaning it can detect objects beyond its own training set and offers greater flexibility to fine-grained audios. In addition, CLIP’s heatmap may offer a better approximation of object size and position than YOLO's bounding boxes.}

\section{Expert Study}
\label{section:expert-study}
To evaluate the usefulness of Soundify in assisting video editors, we conduct a within-subjects expert study comparing Soundify to a baseline task of manual editing. The following outlines our study design, participants, procedure, and results. Our main research questions are:

\begin{itemize}
    \item [RQ1.] How would the level of workload for participants be affected with the use of Soundify?
    \item [RQ2.] How do participants' find the usability of Soundify?
    \item [RQ3.] How would task competition time change with the use of Soundify?
    \item [RQ4.] Qualitatively, what would participants see as the pros and cons of Soundify?
\end{itemize}

\subsection{Study Design}

\subsubsection{Independent Variable}

The independent variable of the study is the system: Soundify versus a baseline of manual editing. In the experimental condition, we ask participants to match sounds to a video using Soundify. Participants may optionally perform creative sound decisions by manually trying out different variations of sound effects. For example, given that Soundify has recommended adding a car sound to a scene, the participant can optionally choose from car sounds of different car models, supplied by the Epidemic Sound library. In the baseline condition, we ask participants to match sounds to a video using Adobe Premiere Pro, a standard video editing software that editors use to add sounds to videos. We provide participants access to the Epidemic Sound library to manually search for audio clips. \rev{Participants manually search for audio clips with keyword search (similar to how they would work today with audio databases).}

\subsubsection{Dependent Variable}

The dependent variables of the study are workload (RQ1) measured by the mental, effort, and frustration components of the NASA TLX questionnaire \cite{nasa-tlx}, usability (RQ2) measured by the System Usability Scale (SUS) questionnaire \cite{sus}, and task completion time (RQ3) reported by the participant (in seconds). All questionnaire questions are represented on a 7-point Likert scale.

\subsection{Participants}

We recruit 12 professional video editors (10 male, 2 female) aged 24 to 59 (mean=34.17, SD=9.81) from Upwork, a platform for hiring freelancers \cite{upwork}. We conduct a background survey with the participants before each study to assess their video and sound editing experience. Overall, participants have high self-rated familiarity with video and sound editing (mean=6.58, SD=0.67) (7-point Likert scale) and have several years of experience editing sound for videos (mean=10.00, SD=7.21). All participants regularly use Adobe Premiere Pro for video and sound editing. In addition, participants also have experience with other video and sound editing software such as Adobe After Effects, Final Cut Pro, DaVinci Resolve, Adobe Audition, Avid Pro Tools, Reaper, and Ableton Live. Furthermore, all participants make use of sound effects libraries in their workflows.

\subsection{Procedure}

We conduct the expert study remotely. After receiving the participant's consent, we collect information about individual backgrounds. We then ask the participant to match sound to a video with Soundify and to match sound to another video with the baseline of manually searching for sounds and adding them in Adobe Premiere Pro. In the experimental condition, the participant may optionally export the results from Soundify and perform further manual editing. We counterbalance both the order of the conditions and the order of the videos. The two videos are of comparable difficulty and contain multiple complex scenes with scenes containing multiple objects. We also ask participants to record the time they spend in each condition. After each condition, we ask participants to complete the NASA TLX, SUS, and task completion time questionnaires. After the participant completes both conditions, we ask the participant to answer open-ended questions regarding the overall experience of using Soundify. The study lasts for approximately 40 minutes. We compensate participants \$30 USD for their time.

\subsection{Results and Discussion}
\begin{figure*}[tbp]
  \centering
  \includegraphics[width=15cm]{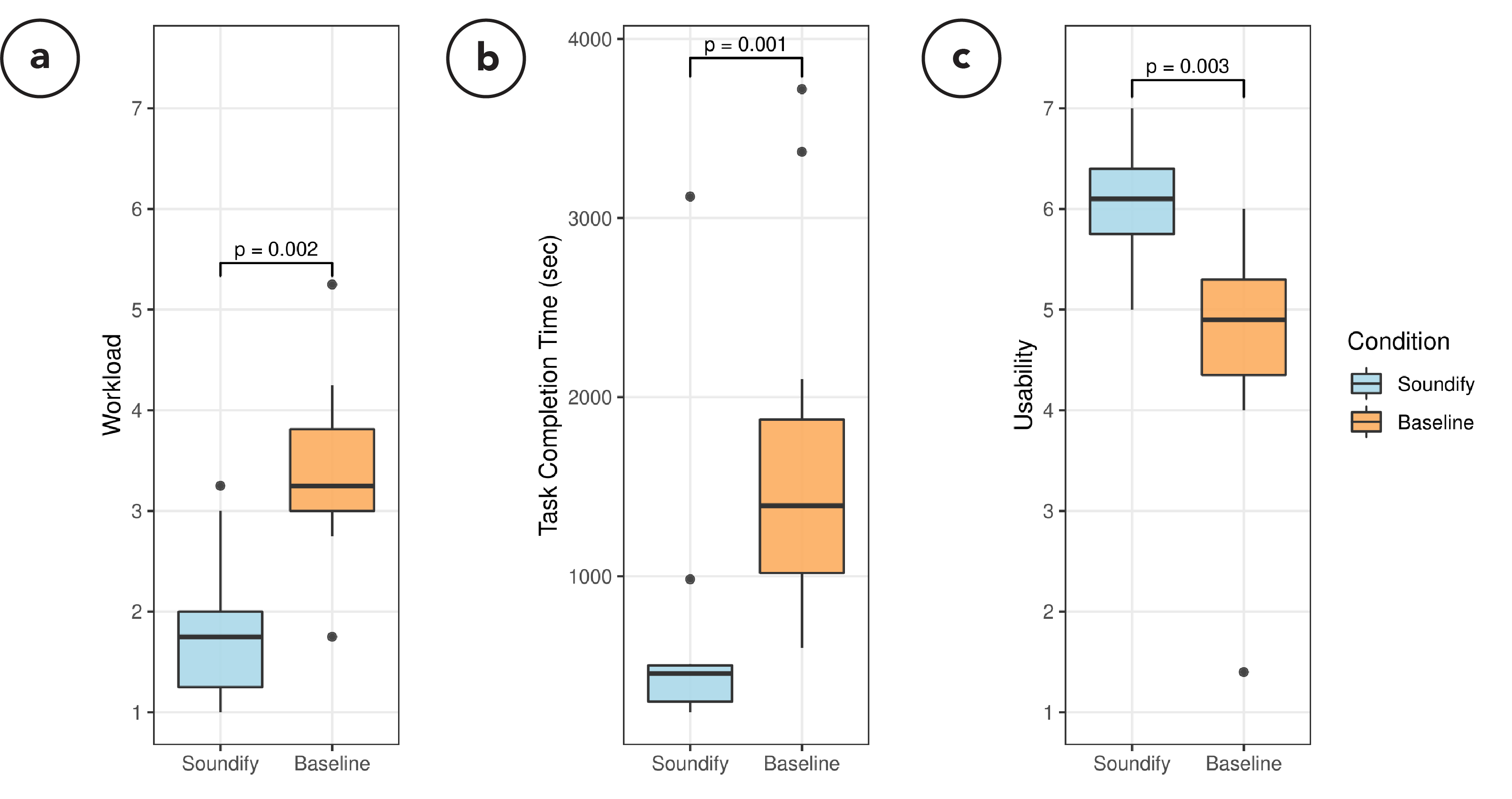}
  \caption{Expert study results ($N$=12). Boxplots from left to right: (a) workload measured with NASA TLX \cite{nasa-tlx} (7-point Likert scale, lower is better), (b) task completion time (seconds, lower is better), and (c) usability measured with SUS \cite{sus} (7-point Likert scale, higher is better).}
  \Description{Expert study results (N=12). Boxplots from left to right: workload measured with NASA TLX (7-point Likert scale, lower is better) (a), task completion time (seconds, lower is better) (b), and usability measured with SUS (7-point Likert scale, higher is better) (c).}
  \label{fig:expert-study-results}
\end{figure*}

For quantitative analysis, we analyze the scores for workload, usability, and task completion using a paired t-test comparing Soundify with the baseline condition. Figure \ref{fig:expert-study-results} shows an overview of the quantitative results comparing Soundify against the manual editing baseline. For qualitative analysis, we analyze the participants' open-ended responses with deductive thematic analysis \cite{thematic-analysis} according to the dimensions of the quantitative measurements (i.e., workload, task completion time, and usability). The following presents and discusses the results of the expert study.

\subsubsection{Workload}
The differences in workload per participant across conditions pass the Shapiro-Wilk test of normality ($W$=0.95, $p$=0.71). We thus compare the differences in workload through a parametric paired t-test. Participants report a significantly lower workload when using Soundify (mean=1.85, SD=0.74) compared to the baseline (mean=3.38, SD=0.87) ($t$(11)=4.02, $p$=0.002, $r$=0.77, $d_s$=1.16) (7-point Likert scale, lower is better) (Figure \ref{fig:expert-study-results}a). Participants express that they "\textit{do edits that feature a lot of tedious sound effect placement constantly} (P5)" and that it's one of their "\textit{least favorite parts of the editing process} (P5)": "I\textit{ feel like most sound effect placement in video editing is essentially grunt work. It doesn't take any kind of skill per se. It's just something you have to do, and it's tedious to find and place these sounds} (P5)". Participants feel that Soundify "\textit{picked appropriate SFX} (P9)" and helped "\textit{minimize having to work on an unfulfilling part of the job} (P5)": "\textit{The AI worked like a charm… everything was with such little effort} (P6)". Participants enjoy "\textit{how little [editing] work was actually needed to be done by the editor} (P10)" and being able to feel more like "\textit{directing and supervising} (P10)" a project: "\textit{Any time I can work in dropdown menus rather than a timeline, I'd prefer it.} (P5)"

\subsubsection{Task Completion Time}
The differences in task completion time per participant across conditions pass the Shapiro-Wilk test of normality ($W$=0.90, $p$=0.14). We thus compare the differences in task completion time through a parametric paired t-test. Participants report a significantly lower task completion time in seconds when using Soundify (mean=670, SD=795) compared to the baseline (mean=1681, SD=979) ($t$(11)=4.40, $p$=0.001, $r$=0.80, $d_s$=1.27) (time taken in seconds, lower is better) (Figure \ref{fig:expert-study-results}b). Participants enjoy "\textit{having a tool that creates a base layer of audio} (P4)" and being able to "\textit{achieve a decent result very quickly} (P12)": "\textit{I love the instantaneous result of being able to instantly bring a scene to life with sound.} (P6)" Participants state that Soundify helps cut down time ("\textit{my time spent was cut in half} (P10)") and could be especially useful when they have a "\textit{high-volume of projects} (P3)": "\textit{I have so many projects to edit and I went from 35 minutes to 7 minutes… I could be so much more efficient in my editing projects if I were to use Soundify in my workflow.} (P8)"

\subsubsection{Usability}
The differences in usability per participant across conditions pass the Shapiro-Wilk test of normality ($W$=0.90, $p$=0.15). We thus compare the differences in usability through a parametric paired t-test. Participants report a significantly higher usability when using Soundify (mean=6.03, SD=0.53) compared to the baseline (mean=4.70, SD=1.23) ($t$(11)=-3.77, $p$=0.003, $r$=0.75, $d_s$=1.09) (7-point Likert scale, higher is better) (Figure \ref{fig:expert-study-results}c). Participants find Soundify to be "\textit{easy to learn} (P3)" and "\textit{easy to use with no instructions} (P9)": "\textit{It was easy to use and accomplished its prescribed goal well.} (P1)" Participants enjoy "\textit{being able to scroll through the portions of the video [scene-by-scene]} (P7)": "\textit{I would use the tool [Soundify] just for the ability to split scenes and file management.} (P11)" Participants mention that the export function is useful: "\textit{I love being about to adjust the audio levels afterward.} (P6)" Participants like how the "\textit{downloaded files are already cut per scene} (P11)" and find that the "\textit{saved numbered format [per scene] is very helpful} (P11)".

Participants also comment on specific technical components of Soundify (classify, sync, and mix):

\textbf{Classify.} Participants state that Soundify "\textit{knocks down the amount of time spent searching for sounds} (P9)": "\textit{I love that it finds the sound effects based on the visuals so you don't have to go keyword treasure hunting for the right sound} (P5)".

\textbf{Sync.} Participants "\textit{appreciate the accuracy with which the sound synchronized to each shot} (P6)": "\textit{I like that it puts the sound in the right place in the timeline for you.} (P5)"

\textbf{Mix.} Participants feel that Soundify "\textit{predicted the panning well} (P7)" and helped "\textit{eliminate the inefficient keyframing within Adobe [Premiere]} (P7)".

\section{Future Work}
\label{section:future-work}

While Soundify was positively received in our user studies, there are several avenues for improvement that we plan to address for future work. 
First, we could improve very fine-grained synchronizations for certain sounds, such as footsteps. \rev{Currently, Soundify does not support matching footsteps to the exact moments in which the foot impacts the ground. Several potential approaches for exploration may include incorporating motion cues, conducting state analysis, and leveraging the metadata of videos (e.g., timecoded script with sound annotations).}
Second, CLIP (the base model that Soundify is built on) may occasionally encounter mistakes. One example is CLIP incorrectly classifying an AC repair handyman holding a screwdriver as a person brushing their teeth. To ensure the accurate classification of sound objects at the beginning of the Soundify pipeline, we will continually update Soundify with the latest improved CLIP model \cite{clip-improve}. \rev{In addition, a potential avenue for exploration may be to finetune CLIP on a large set of labels found in the audio library.}
Third, video editors also suggested several manual finetuning capabilities, such as custom fades and effects and adjusting EQ (i.e., boosting or damping certain frequencies). We could allow the tuning of additional effects in future work by integrating Soundify into a more comprehensive sound editing software.

\section{Conclusion}
We identify the challenge of video editors in manually adding sounds to video through a formative interview ($N$=10) and distill a set of principles to guide the development of our solution. In this paper, we present Soundify, a system that assists editors in matching sounds to video that works out-of-the-box for arbitrary audio categories. Our key insight is a combination of leveraging studio-quality sound effects libraries and repurposing CLIP, an image classification model, into a sound localizer. Given video footage, Soundify assists the video editor in surfacing relevant sound clips, synchronizing the sound clips to the video in time, and converting the sound clip to spatial audio by dynamically adjusting the sound clip's panning and volume through exploiting CLIP's activation maps. We conduct a human evaluation study ($N$=889 raters), evaluating Soundify's results in automatically matching sounds to video for a diverse range of audios. We further evaluate the usefulness of Soundify for video editors through a within-subjects expert study ($N$=12) comparing Soundify to a baseline (manual editing with Adobe Premiere Pro), showing decreased workload, decreased task completion time, and increased usability.


\begin{acks}
\rev{We would like to thank all our user study participants, peers in our research group for providing valuable feedback on iterations of our system and paper write-up, and the anonymous reviewers for their comments.}
\end{acks}

\bibliographystyle{ACM-Reference-Format}
\bibliography{base}
\pagebreak

\appendix

\newpage
\onecolumn

\section{Detecting multiple objects}

\begin{figure*}[tbp]
  \centering
  \includegraphics[width=14cm]{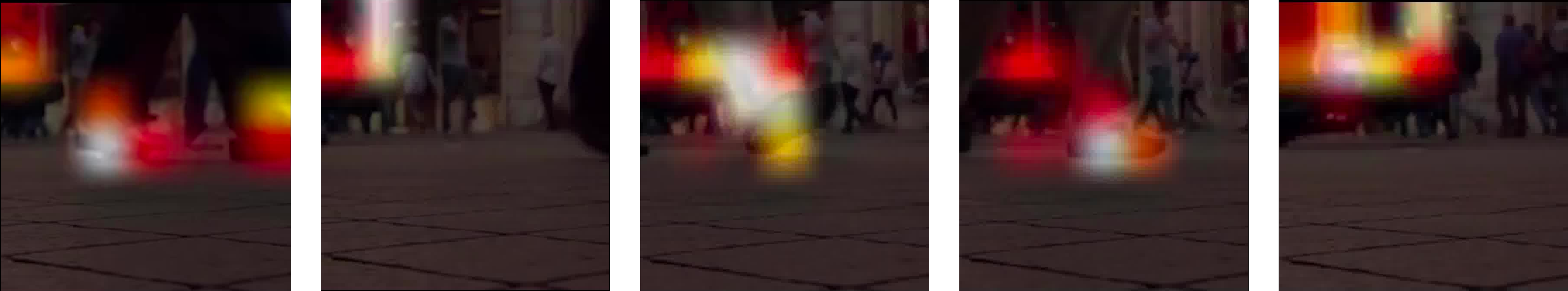}
  \caption{Example frames of multiple objects (footsteps + tram) being detected.}
  \Description{Example frames of multiple objects (footsteps + tram) being detected.}
  \label{fig:multiple-footsteps-tram}
\end{figure*}

\rev{Soundify can detect multiple sound objects simultaneously, such as both footsteps and tram in Figure \ref{fig:multiple-footsteps-tram}. We provide some more examples in the paper’s project page. For non-salient objects, Soundify can capture their sounds if they are visible on screen (e.g., water ripple). We provide an ambient sound classification to capture ambient sounds of objects that may not be visible on screen (see \ref{classify}).}

\end{document}